\documentstyle[12pt]{article}
\input epsf.tex

\newcommand{\bmat}{\left(\begin{array}}
\newcommand{\emat}{\end{array}\right)}
\def\NPB#1#2#3{Nucl. Phys. B{#1} (19#2) #3}
\def\PLB#1#2#3{Phys. Lett. B{#1} (19#2) #3}

\def\PRD#1#2#3{Phys. Rev. D{#1} (19#2) #3}

\def\yzero{\smash{\hbox{$y\kern-4pt\raise1pt\hbox{${}^\circ$}$}}}
\def\ov{\overline}
\def\s2{\frac{1}{\sqrt2}}

\def\beq{\begin{equation}}
\def\eeq{\end{equation}}
\def\beqa{\begin{eqnarray}}
\def\eeqa{\end{eqnarray}}

\def\diag{{\rm diag \,}}

\def\Dsl{\,\raise.15ex\hbox{/}\mkern-13.5mu D} 
\def\IR{\bf R}
\def\IC{\bf C}
\def\IZ{\bf Z}
\def\ItZ{\bf {\tilde Z}}

\def\IT{{\bf T}}
\def\IS{{\bf S}}
\def\IRP{{\bf RP}_2}

\def\IXAH{{\bf X_{AH}}}

\def\id{{\bf 1}}
\def\Dseven{${\widehat{{\rm D}7}}$}
\def\Deight{${\widehat{{\rm D}8}}$}
\def\Dzero{${\widehat{{\rm D}0}}$}
\def\Dthree{${\widehat{{\rm D}3}}$}
\def\Dfour{${\widehat{{\rm D}4}}$}
\def\Dfive{${\widehat{{\rm D}5}}$}

\def\Dbnine{${\ov{{\rm D}9}}$}
\def\Dbseven{${\ov{{\rm D}7}}$}

\def\Dbtwo{${\ov{{\rm D}2}}$}

\def\Dbp{${\ov{{\rm D}p}}$}

\topmargin
-1.5cm
\textwidth
15.5cm
\textheight
24cm
\oddsidemargin
0.7cm
\evensidemargin
1.2cm

\begin{document}

\makeatletter
\@addtoreset{equation}{section}
\makeatother
\renewcommand{\theequation}{\thesection.\arabic{equation}}
\pagestyle{empty}
\rightline{CERN-TH/2001-094}
\rightline{CINVESTAV-IPN/25-01}

\rightline{\tt hep-th/0104173}
\vspace{2.5cm}
\begin{center}
\Large{\bf The fate of the type I non-BPS D7-brane}\\

{\large
Oscar Loaiza-Brito \footnote{On leave from Departamento de F\'{\i}sica,
Centro de Investigaci\'on y Estudios Avanzados del IPN (CINVESTAV-IPN),
M\'exico D.F., M\'exico, e-mail: oloaiza@fis.cinvestav.mx}, 
Angel~M.~Uranga \\[2mm]
{\em Theory Division, CERN}\\
{\em CH-1211 Geneva 23, Switzerland} \\[4mm]}

\vspace*{1.5cm}

\small{\bf Abstract} \\[7mm]
\end{center}

\begin{center} \begin{minipage}[h]{14.0cm}
{\small
We describe the fate of the Type I non-BPS D7-brane, which is tachyonic
but carries a non-trivial K-theory $\IZ_2$ charge. It decays to topologically 
non-trivial gauge field configurations on the background D9-branes. In the 
uncompactified theory the decay proceeds to infinity, while with a transverse 
torus the decay reaches a final state, a toron gauge configuration with
vanishing Chern classes but non-trivial $\IZ_2$ charge. A similar behaviour 
is obtained for the type I non-BPS D8-brane, and other related systems. 
We construct explicit examples of type IIB orientifolds with non-BPS
D7-branes, which are hence non-supersymmetric, but for which supersymmetry
is restored upon condensation of the tachyon.

We also report on the interesting structure of non-BPS states of type IIA
theory in the presence of an O6-plane, their M-theory lifts, the
relation between string theory K-theory and M-theory cohomology, and its
interplay with NS-NS charged objects. We discuss several new effects,
including: i) transmutation between NS-NS and RR torsion charges, ii) 
non-BPS states classified by K-theory but not by cohomology in string
theory, but whose lift to M-theory is cohomological.
}

\end{minipage}
\end{center}

\bigskip

\bigskip

\leftline{CERN-TH/2000-094}

\leftline{April 2001}

\newpage
\setcounter{page}{1}
\pagestyle{plain}
\renewcommand{\thefootnote}{\arabic{footnote}}
\setcounter{footnote}{0}

\section{Introduction}

Type I string theory contains certain D-branes with conserved K-theory
charges, but which are nevertheless tachyonic. Specifically, type I non-BPS 
\Dseven- and \Deight-branes \footnote{We use a bar to denote antibranes,
and a hat to denote non-BPS branes} carry non-trivial K-theory $\IZ_2$
charges, but contain world-volume tachyons arising from strings stretching
between the non-BPS brane and the background D9-branes \cite{lerda}. Given
their non-trivial conserved charges, such states must decay not to the
vacuum, but to some other state with their same quantum numbers. In this
note we give a detailed picture of this decay, intending to clarify some
confusion in the literature. We would like to mention however that some of
the points have been anticipated in \cite{sugi, bergman}.

Intuitively, one expects these branes to decay to gauge field configurations 
on the D9-brane world-volume, associated to topologically non-trivial 
bundles in the correct K-theory class. In fact, in Section 2 we argue that
the \Dseven-brane in the uncompactified theory is unstable against
dissolving as a monopole-antimonopole pair in the D9-brane gauge group. The 
tachyon has a runaway behaviour and does not reach a minimum for any
finite characteristic size of the gauge field lump. Hence the decay
continues to infinity, leading to an infinitely extended and diluted gauge
configuration. If the space transverse to the \Dseven-brane is
compactified on a two-torus, the decay reaches a final state, which we
characterize in detail as a  $\IZ_2$ toron of $SO(32)$ gauge theory. We
also describe a T-dual picture in which the decay process is very
intuitive, and corresponds to recombining intersecting D-branes into
smooth ones, in the spirit of e.g. \cite{afiru,blumbound}.

In Section 3 we analyze the decay of the \Deight-brane, whose story is
similar. It decays to a `kink' configuration in the D9-brane gauge
group, which in a non-compact setup does not reach any final state but
becomes infinitely extended and diluted. In compact space it reaches a
final state described by a $\IZ_2$ Wilson line on the D9-branes.
In Section 4 we describe related systems of D-branes and O-planes. 

In Section 5 we extend on an independent topic, originally motivated by
the above systems. We consider the system of an O6-plane with no
overlapping D6-branes. By the previous analysis this theory contains
stable non-BPS \Dfour- and \Dfive-branes, carrying $\IZ_2$ charges. In
discussing the M-theory lift of these and other stable non-BPS states in
the configuration, we find multiple interesting issues concerning the
nature of NS-NS and RR charges in the IIA theory, and its M-theory origin.
In particular, we find that certain non-BPS states carrying torsion NS-NS
and RR charges are topologically equivalent once lifted to M-theory. We
also find non-BPS states in IIA theory whose charge is classified by
K-theory but not by cohomology, but whose M-theory lift corresponds to
M-branes wrapped on torsion cohomology cycles in the Atiyah-Hitchin
manifold. This result is relevant in understanding the appearance of
K-theory in string theory as derived from M-theory, and we explain why
it suggests an extension of those in \cite{dmw}. This section is
self-contained, and the reader interested only in these aspects is adviced
to proceed directly to it.

Finally, in the appendix we construct an explicit type IIB orientifold
with \Dseven-branes, where a proper understanding of the tachyon is quite
essential, since supersymmetry is restored upon condensation of the
\Dseven-brane tachyon.

\section{The fate of the type I non-BPS D7-brane}

\subsection{Construction}

As described in \cite{wittenkth}, the type I non-BPS \Dseven-brane is
constructed as a pair of one D7- and one \Dbseven-brane in type IIB,
exchanged by the action of world-sheet parity $\Omega$. In the
world-volume of the $7{\ov 7}$ pair, before the $\Omega$ projection, the
8d spectrum is as follows: In the 77 sector we have a $U(1)$ gauge
boson, one complex scalar, and fermions in the $4+{\ov 4}$ of the $SO(6)$
Lorentz little group; the ${\ov 7}{\ov 7}$ sector leads to an analogous
piece; the $7{\ov 7}+{\ov 7}7$ sector provides one complex tachyon and two
fermions in the $4+{\ov 4}$, all with with $U(1)^2$ charge $(+1,-1)$. The
$\Omega$ action exchanges the $77$ and ${\ov 7}{\ov 7}$ sectors, leaving a
group $U(1)$, a complex scalar and vector-like fermions, and maps the
$7{\ov 7}+{\ov 7}7$ to itself, projecting out the tachyon and keeping just
one set of fermions.

Hence the type I \Dseven-brane would appear to be tachyon-free. However,
RR tadpole cancellation in type I requires the presence of a background of
32 D9-branes, the role of which for the stability of the \Dseven-brane
was noticed in \cite{lerda} (see also \cite{schwarz}). In fact, the
$79+97$ sector gives a complex tachyon field, and one chiral fermion,
while the ${\ov 7}9+9{\ov 7}$ sector gives their $\Omega$ image.

Despite the instability associated to this tachyon, the D7-brane is unable
to decay to the vacuum. This follows from the fact that it carries a 
$\IZ_2$ charge corresponding to the non-trivial K-theory class $x$ in
${\bf KO}(\IS^2)=\IZ_2$. In fact, as proposed in \cite{wittenkth} and
checked in \cite{gukov}, the $\IZ_2$ charge can be detected by a $-1$ sign
picked up by a \Dzero-brane probe moving around the \Dseven-brane in the
transverse two-plane. Since the \Dzero-brane transforms as a $SO(32)$
spinor \cite{senspinor}, this implies that the K-theory class associated
to the \Dseven-brane corresponds to a topologically non-trivial bundle
with asymptotic monodromy in the non-trivial element of 
$\Pi_1(SO(32))=\IZ_2$. This property in fact characterizes the non-trivial
class $x$ in ${\bf KO}(\IS^2)=\IZ_2$

A different way to detect the $\IZ_2$ charge of the \Dseven-brane  
is to introduce a D5-brane probe intersecting the \Dseven-brane over a
four-dimensional space \cite{probe}. The intersection leads to a single
four-dimensional Weyl fermion doublet of the D5-brane $SU(2)$ gauge group. 
The $\IZ_2$ global gauge anomaly \cite{witten} from this 4d fermion is a
reflection of the non-trivial $\IZ_2$ charge of the \Dseven-brane
\footnote{In the non-compact context, this anomaly is presumably
cancelled by a (K-theoretic) anomaly inflow mechanism, implicit in
\cite{freed}. Hence there is no inconsistency in the configuration,
corresponding to the fact that D-brane charge need not cancel in
non-compact space.}. In fact, the appearance of such fermion implies that
the K-theory class of the \Dseven-brane corresponds to a bundle with an
odd number of fermion zero modes of the (real) Dirac operator. By the
index theorem \cite{atisin}, this characterizes the K-theory class of the
\Dseven-brane as the non-trivial class $x$ in ${\bf KO}(\IS^2)=\IZ_2$.

Hence the type I \Dseven-brane carries a topological charge, endowing it
with quantum numbers different from those of the vacuum, to which it cannot 
decay. It must instead decay to some state carrying the same $\IZ_2$ (and
no other) charge. Since, from the viewpoint of the parent type IIB theory,
the tachyon triggering the decay arises from open strings stretching
between the D9- and D7-branes (the ${\ov 7}9+9{\ov 7}$ sector giving
merely its $\Omega$ image), one is led to suspect that the decay will be
roughly speaking a $\Omega$-invariant version of the much studied tachyon
condensation in the D$p$-D$(p+2)$ system (see e.g. \cite{minustwo}). In
fact, our analysis below will make this analogy quite precise, and
establish that the \Dseven-brane dissolves as a non-trivial gauge field
configuration on the D9-brane gauge group. Much as in the D$p$-D$(p+2)$
system, for non-compact transverse space the tachyon condensation does not
reach an endpoint and leads to infinitely extended and diluted gauge
configurations. For compact transverse space, however, the tachyon reaches
a minimum and the condensation reaches an endpoint configuration.

\subsection{The non-compact case}

We have argued that the \Dseven-brane is unstable against decay to a
topologically non-trivial gauge bundle on the D9-branes, characterized by
the K-theory class $x$ associated to the \Dseven-brane $\IZ_2$ charge.
Indeed, it is easy to describe $SO(32)$ gauge configurations on $\IR^2$
carrying such charge. Before going into details, let us emphasize that
such configurations are however not solutions of the equations of
motion. A scaling argument \cite{wittenkth} shows that any such lump
configuration can lower its energy by increasing its characteristic
size. Hence, the gauge field lumps tend dynamically to become infinitely
extended and diluted \footnote{The absence of a tachyon vacuum manifold
allows to avoid the argument in \cite{schwarz}. For more detailed
discussion of the relation, in this example, between the classification of
states by K-theory classes vs. homotopy classes of the vacuum manifold,
see \cite{bergman}.}. However, it is interesting to explore the properties
of the topological class of such configurations.

In order to describe a simple example, let us split the group $SO(32)$ as
$SO(30)\times SO(2)$, and embed a non-trivial gauge background on the
$SO(2)$ factor, of the form
\beqa
F= f(x_1,x_2) \pmatrix{0 & 1 \cr -1 & 0}
\label{lump}
\eeqa 
Here $f(x_1,x_2)$ is a function with compact support in a subset $\Sigma$
or $\IR^2$. 

We should require proper Dirac quantization for fields in the adjoint of
$SO(32)$. This representation decomposes as ${\bf 496}={\bf 435}_0+
{\bf 30}_{+1}+{\bf 30}_{-1}+{\bf 1}_0$, with subindices denoting the
$SO(2)$ charge. Hence, Dirac quantization for this representation requires
the integral of $f$ to be $2\pi$ times an integer. Choosing the minimum
Dirac quantum
\beqa
\int_\Sigma f(x_1,x_2) dx_1 dx_2 = 2 \pi,  
\label{quantum}
\eeqa
quantization is obeyed for the $SO(32)$ adjoint, but not for the $SO(32)$
spinor representation. Fields in this representation carry charge 
$q=\pm 1/2$ under the $SO(2)$ subgroup, hence their asymptotic holonomy is
\beqa
\exp (q\oint_{\partial\Sigma} A ) = \exp (q\int_\Sigma F) = \exp(q\,2\pi i
\sigma_2 ) = -\id_2
\eeqa
Hence \Dzero-branes pick up a $-1$ phase in going around the gauge field
lump (\ref{lump}), which therefore carries the correct K-theory $\IZ_2$ 
charge.  Notice that, regarding the $SO(2)$ subgroup as arising from two
D9-branes (related by $\Omega$), the above gauge background can be seen as
a unit Dirac monopole in one D9-brane (we denote it D9$_+$) and an
antimonopole in another (D9$_-$), both being exchanged by $\Omega$.

It is also easy to detect the $\IZ_2$ topological charge by introducing a
D5-brane probe, spanning the directions $x^1$, $x^2$. The 4d Weyl fermions
in doublets of the D5-brane $SU(2)$ arise from zero modes in the
Kaluza-Klein reduction of the 6d fermions in the 59+95 sector, in the
presence of the gauge field background. Clearly, strings stretched between
D5-branes and the 30 D9-branes with no gauge flux lead to no contribution
to the 4d global gauge anomaly. However, fermions zero modes arise from
strings stretched between the D5-branes and the D9$_+$-brane (while strings 
stretched between the D5- and the D9$_-$-brane are merely their $\Omega$
image). The number of fermion zero modes is given by the index of the
(complex) Dirac operator coupled to the monopole background. By the index
theorem, this number is $1$ for the minimum quantum (\ref{quantum}),
reproducing the correct 4d Weyl fermion global anomaly. The above
discussion can be regarded as a rudimentary computation of the mod two
index of the real Dirac operator of the $SO(32)$ gauge configuration.

Notice that for any such configuration, strings stretched between the 30
D9-branes with no background, and the D9$_+$- or the D9$_-$-brane lead to
additional tachyons, whose condensation would further break the gauge
group. Such strings behave as charged particles in a magnetic field, and
so are localized in $\IR^2$. The negative squared masses of the tachyons
are related to the gauge field strength at such points. Hence, as the flux
becomes more dilute the masses get less and less tachyonic. This argument
also supports that the system become more and more stable as the gauge
lump expands.

\subsection{Toroidal transverse space}

The situation would be better understood by taking the space transverse
to the \Dseven-brane to be a $\IT^2$, since the size of the gauge field
lump would have an upper bound, and we may expect a well-defined final
state.  Notice that there is a subtlety in trying to do so: As
discussed in \cite{probe}, consistency of the configuration requires
cancellation of the K-theory $\IZ_2$ charge in the compact space, so we
have to include at least two \Dseven-branes. The total configuration hence
has the same quantum numbers as the vacuum, to which it could annihilate. 
However, for our present purposes we may consider one of the \Dseven-branes
as spectator, and ignore it in the analysis of the decay of the other. We
will be interested in describing the decay in this setup \footnote{A more
suitable system to address the compact setup would be the (isomorphic in
other respects) decay of a non-BPS D$(p-2)$-brane in the presence of
D$p$-branes on top of an O$p$-plane, See section 4. The existence of this
non-BPS brane, its non-trivial $\IZ_2$ charge, and its tachyon in the
$p$-$(p-2)$ sector, are shown in the same way as for the \Dseven-brane. For 
lower $p$ we may consistently compactify without bothering about tadpole
cancellation conditions.}.

\medskip

\subsubsection{A T-dual picture}

\begin{figure}
\begin{center}
\centering
\epsfysize=6cm
\leavevmode
\epsfbox{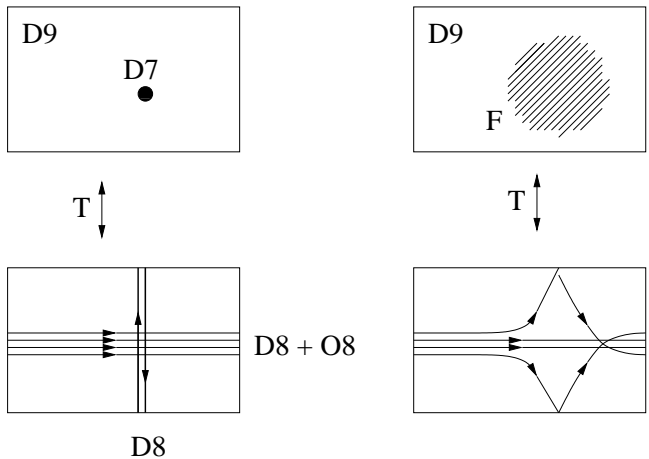}
\end{center}  
\caption[]{\small Schematic picture of the decay of the type I
\Dseven-brane in the type I and T-dual type I' pictures}
\label{fig1}   
\end{figure}

One advantage in making the transverse space compact is the possibility of
using T-dual pictures to describe the decay process. Considering for
simplicity a square two-torus, and vanishing NS-NS B-field, we may
T-dualize along one direction in the torus to get type I' theory, as
depicted in Fig~\ref{fig1}. The O9-plane becomes two O8-planes, and the
D9-branes become D8-branes, all depicted horizontally in the picture. The
type IIB D7-${\ov {{\rm D}7}}$ pair becomes a pair of oppositely oriented
vertical D8-branes, which are exchanged by the orientifold action $\Omega
R$, where $R$ is a reflection in the vertical direction.  Some useful
references for the T-duality relation and what follows are
\cite{bgkl,afiru,blumbound}.

The process of dissolving the D7-${\ov{{\rm D}7}}$ pair in the D9-branes
corresponds in the T-dual to recombining the intersecting vertical and
horizontal D8-branes into a smooth one, as shown in the figure (see e.g.
\cite{afiru,blumbound}). The gauge group on the D8-branes is broken to a
subgroup of $SO(32)$, corresponding to the fact that the magnetic flux
breaks the D9-brane group to a subgroup. In particular the above $SO(2)$
monopole-antimonopole background corresponds to recombining two horizontal
D8-branes with the vertical ones, in a $\Omega R$ invariant fashion.

This picture makes it clear that, in the non-compact context (large volume
limit of the original torus, large complex structure of the T-dual torus),
the process of recombination does not reach a final state, but goes on to   
infinity, since it is always possible to lower the length of the 
recombined D-branes by straightening them.

The $\IZ_2$ charge carried by the configuration is not so manifest in
terms of moving around (the T-dual of) a \Dzero-brane. However, it is
easily detected by introducing the T-dual of a D5-brane probe, which is a
pair of horizontal D4-branes. The 4d Weyl fermion arises from the
intersection between the D4-branes and the recombined D8-branes.

Finally, this picture makes manifest the evolution of the spectrum of
the configuration as the recombination proceeds. In particular, the
tachyons arising from the remaining intersections have negative
squared masses proportional to the intersection angles. Hence, it is
obvious they become less and less tachyonic as the configuration relaxes
by extending and diluting, as discussed above.

\medskip

\subsubsection{The final configuration}

In the compact context the decay process reaches a final state 
(equivalently the tachyon potential has a minimum), whose description and
properties we study in this section.

In the type I' picture it is clear that any intersection implies the
existence of a tachyon triggering recombination. The system does not
stabilize until all branes are aligned in the horizontal direction. This
agrees with that configuration being the minimum energy state with total
homology class $(32,0)$ in $\IT^2$. The original $\IZ_2$ charge is not
homological, but rather K-theoretical, and not too manifest in this
picture. Because of this we describe the final state in the original type
I picture.

In terms of the original picture, the final state is a D9-brane gauge
field configuration, with vanishing Chern classes, and with a topological 
$\IZ_2$ charge, characterized by $-1$ holonomies in the spinor 
representation. There is a simple gauge field configuration satisfying
those requirements, a $SO(32)$ toron.

Consider decomposing $SO(32)$ as $SO(29)\times SO(3)$. Since $SO(3)\simeq 
SU(2)/\IZ_2$, one can construct a set of two Wilson lines which commute in
$SO(3)$ but anticommute in $SU(2)$. In the two-dimensional
representation of $SU(2)$ we can take
\beqa
P=\pmatrix{i & 0 \cr 0 & -i} \quad ; \quad Q=\pmatrix{0 & i \cr i & 0}
\eeqa
which satisfy $PQ=-QP$. In the three-dimensional representation, the
vector of $SO(3)$, these matrices descend to the (commuting) matrices
\beqa
\gamma_1=\diag (1,-1,-1)  \quad ; \quad \gamma_2=\diag(-1,1,-1)
\label{triplet}
\eeqa
This construction is equivalent to that of $SU(2)$ toron configurations in
\cite{torons}.

We propose this configuration as the final state for the decay of the type
I \Dseven-brane. In fact, the connection is flat and all Chern classes
vanish. However, the corresponding bundle is non-trivial, and  carries the
correct $\IZ_2$ charge, as we now discuss \footnote{The $\IZ_2$ charge of
the Wilson lines (\ref{triplet}) was already noticed in \cite{sugi}.}.

For instance, a \Dzero-brane transforms as a $SO(32)$ spinor, hence as a
doublet under the above $SU(2)$. In the presence of such gauge background,
it suffers a holonomy $PQP^{-1}Q^{-1}=-1$ when carried along the square
surrounding the unit cell in $\IT^2$. This characterizes the existence
of the correct $\IZ_2$ charge in the configuration.

It is also easy to detect the $\IZ_2$ charge by introducing a D5-brane
probe wrapped on $\IT^2$. The spectrum in the 59+95 spectrum, in the
absence of gauge background, consists of half-hypermultiplets in the
representation $({\bf 2};{\bf 32})$, i.e. an even number of doublets under
the D5-brane $SU(2)$. In the presence of Wilson lines, they decompose as
half-hypermultiplets in the $({\bf 2};{\bf 29},{\bf 1})+({\bf 2};{\bf 1},
{\bf 3})$. The former are untouched, but the latter are projected out by
the matrices (\ref{triplet}). We are hence left with an odd number of Weyl
fermions doublets of $SU(2)$, whose contribution to the global gauge
anomaly reflects the non-trivial $\IZ_2$ charge of the gauge bundle.

Let us conclude this section by pointing out that the final state is
unique, up to gauge transformations. Another interesting property is that
the final state in the decay of the \Dseven-brane is supersymmetric, and
obviously free of tachyons. This is consistent with its being the minimum
energy configuration in its topological class. This remark is important in
certain applications, like model building. This is illustrated in the
appendix, where we construct a type IIB non-supersymmetric orientifold
with \Dseven-branes, with supersymmetry restored upon tachyon
condensation.

\section{The fate of the type I non-BPS D8-brane}

In this section we would like to discuss the decay of the \Deight-brane. 
Since it is analogous to the decay of the \Dseven-brane in some respects,
our treatment is more sketchy.

The type I \Deight-brane can be described in open string theory by 
imposing Dirichlet boundary conditions in one direction, say $x^9$, and
endowing worldsheet boundaries with an additional fermion degree of
freedom \cite{wittenkth}. In the 88 sector, after the $\Omega$ projection
one obtains a $\IZ_2$ ($\simeq O(1)$) gauge group, one real scalar field,
and one 9d fermion. In the 89+98 sector, one obtains a real tachyon and
one 9d fermion.

Hence, the \Deight-brane is unstable against condensation of the 89
tachyon. However, the state carries a non-trivial K-theory $\IZ_2$ charge,
and cannot decay to the vacuum. The topological $\IZ_2$ charge can be
detected, as proposed in \cite{wittenkth} and checked in \cite{gukov}, by
moving a non-BPS D-instanton across the \Deight-brane, since its amplitude
picks up a $-1$ sign in the process. This characterizes the K-theory class
of the \Deight-brane to be the non-trivial class $y$ in 
${\bf KO}(\IS^1)=\IZ_2$.

An alternative way to detect the $\IZ_2$ charge is to consider a stack 
of $n$ type I D1-branes wrapped along $x^9$. The intersection between the
D1- and the \Deight-branes supports one 1d fermion, leading to a $\IZ_2$
global gauge anomaly in the $O(n)$ gauge group. The existence of this
fermion zero mode characterizes the K-theory class of the \Deight-brane as
the class $y$ in ${\bf KO}(\IS^1)=\IZ_2$.

\medskip

The \Deight-brane therefore decays not to the vacuum, but to a non-trivial
gauge field configuration on the D9-branes, with non-trivial profile only
along $x^9$.  By a scaling argument, such gauge field lumps are not static, 
but dynamically tend to become infinitely extended and diluted. The
tachyon hence does not have a minimum, but rather a runaway behaviour. 
However, it is interesting to consider intermediate steps in its
condensation.

The gauge configurations in this class are characterized by being pure
gauge in the two asymptotic regions $x^9\to \pm \infty$, $A=gdg^{-1}$, with 
$g(x^9\to\pm\infty)$ in different connected components of the perturbative
D9-brane gauge group $O(32)$. The gauge field at infinity is associated to
the non-trivial class in $\Pi_0(O(32))$. By using the spectral flow of the
Dirac operator and the index theorem \cite{aps}, this property ensures
that the type I D-instanton amplitude picks up a $-1$ factor in crossing
the gauge field lump (see \cite{gukov} for details). The $\IZ_2$ charge
can also be made manifest by introducing a D1-brane stretched along
$x^9$. By an argument isomorphic to that in the first reference of
\cite{senspinor}, or using the index theorem, there is generically one
normalizable fermion zero mode in the 19+91 sector, leading to one 1d
fermion and hence to a $\IZ_2$ global gauge anomaly. 

\medskip

The decay is more tractable by compactifying the transverse space on a
circle $\IS^1$. Again, this is possible only if one considers an even
number of \Deight-branes, all of which but one are considered as
spectators in what follows. (Alternatively, one may consider the
(isomorphic is other respects) decay of a non-BPS D$(p-1)$-brane in the
presence of an O$p$-plane and D$p$-branes, see Section 4).

In such configuration, the decay of the \Deight-brane does reach a final
state, which is simply a D9-brane gauge field configuration given by a
non-trivial $O(32)$ $\IZ_2$ Wilson line, of the form 
\beqa
\gamma=\diag(-1,\id_{31})
\eeqa
It is straightforward to check that this configuration carries the correct
$\IZ_2$ charge \footnote{The $\IZ_2$ charge of this Wilson line was observed 
in \cite{sugi}.}. The spectrum of strings stretching between a D-instanton
and the D9-branes contains fermion degrees of freedom in the vector of
$O(32)$. Hence, in going around the circle, the D-instanton amplitude
picks up a $-1$ factor, reflecting the $\IZ_2$ charge carried by the gauge 
configuration. Similarly, if we introduce a D1-brane wrapped along the
$\IS^1$, the 19+91 spectrum contains an odd number of one-dimensional
fermions, leading to a 1d global gauge anomaly which reflects the
background $\IZ_2$ charge.

Hence, very much as for the \Dseven-brane, the decay of the \Deight-brane
proceeds to infinity in the non-compact configuration, but reaches a
supersymmetric final state in the compact case, described as a simple
D9-brane gauge background.

\section{Some related systems}

\subsection{Lower-dimensional O$p$-planes}

A system analogous to the above ones is that of lower-dimensional
non-BPS ${\widehat{\rm{D}q}}$-branes embedded within a set of D$p$-branes,
on top of a negatively charged O$p$-plane
\footnote{The system may contain additional non-BPS states associated to
branes extending away from the O$p$-plane. We do not study them in this
section, but will play a role in Section 5.3.}. The analysis of the open
string spectrum is isomorphic to the above, the only difference being that
the number of D$p$-branes in unconstrained. For configurations with a
non-vanishing number of D$p$-branes there exist non-BPS D$(p-1)$- and
D$(p-2)$-branes which carry a topological $\IZ_2$ charge, but contain
tachyons in the mixed sector. 

As above, one is led to propose that the D$q$-brane dissolves into a gauge
field configuration on the orthogonal gauge group on the D$p$-branes,
which dynamically becomes infinitely extended and diluted. In fact this is
correct in general, save for an interesting subtlety for configuration
with too few D$p$-branes. 

Consider for instance a D$(p-2)$-brane in presence of a single D$p$-brane
on top of an O$p$-plane. There is a $p$-$(p-2)$ tachyon triggering a decay
of the D$(p-2)$-brane to a gauge background. However, a single D$p$-brane
is unable to carry a bundle with the required $\IZ_2$ charge. The way out
of the paradox is to realize that the topological class of the configuration 
is defined up to brane-antibrane nucleation/annihilation. The system hence
nucleates additional D$p$-${\ov{{\rm D}p}}$ pairs \footnote{In general,
it is allowed to nucleate higher-dimensional branes extending away from
the O$p$-plane. To simplify the discussion we restrict to 
D$p$-${\ov{{\rm D}p}}$ pairs.}, leading to an enhanced
gauge group $SO(n+1)\times SO(n)$. Keeping the gauge bundle on the \Dbp-branes 
trivial for simplicity, we may now embed a gauge configuration on the
D$p$-brane group, carrying the correct $\IZ_2$ charge.

The system presents an interesting energetic balance. Nucleation of
D$p$-${\ov{{\rm D}p}}$ pairs is energetically convenient as long as the
$p\bar{p}$ tachyons lie close to their minimum, namely the bundles on
branes and antibranes do not differ much. This prevents the gauge field
lump to become infinitely extended and diluted, since this would lead to a
too large uncancelled D$p$-${\ov{{\rm D}p}}$ pair tension. Instead, the
gauge field lump stabilizes at a finite size, which may be estimated as
follows. Consider compactifying the two DN directions in the problem in a
square $\IT^2$, with equal radii $R$, and nucleate e.g. two D$p$-${\ov{{\rm
D}p}}$ branes wrapped on it. For any $R$, a point-like D$(p-2)$-brane is
tachyonic and tends to dissolve in the D$p$-branes. The opposite situation, 
with a uniform gauge lump, given by a a SO(3) $\IZ_2$-toron, is free of
$p$-${\ov p}$ tachyons for small enough $R$, due to the effect of the
Wilson lines. However, for $R$ larger than the critical value $R_c=
\sqrt{2\alpha'}$, a tachyon develops, suggesting that a uniformly extended 
lump of size larger than $R_c$ is not stable. We take this as
evidence for the stability of a configuration of branes and antibranes,
with a gauge lump of characteristic size $\sqrt{2\alpha'}$, as a result of
two competing effects: decreasing the lump size results in energetic cost
for the lump itself, while increasing its size results in energetic cost
from the brane-antibrane tensions.

\subsection{The $USp(32)$ theory}

As another interesting system of D-branes and O-planes, we consider the
possible existence of unstable but non-trivially charged branes in $USp(32)$
type I theory \cite{sugimoto}. Recall this theory is obtained by modding
out type IIB theory by $\Omega$, but with an O9-plane with positive RR
charge. The RR tadpole is subsequently cancelled by introducing 32
\Dbnine-branes, leading to a gauge group $USp(32)$. D-brane charges are
classified by ${\bf KSp}({\bf S}^n)$ ($\simeq {\bf KO}({\bf S}^{n+4})$
from Bott periodicity), and lead to $\IZ$-valued charges for D1, D5 and
D9-branes, and $\IZ_2$-valued charges for \Dthree, \Dfour-branes. The
latter are easily constructed in string theory (see e.g. \cite{gukov}): the 
\Dthree-brane is obtained as a type IIB D3-${\ov{{\rm D}3}}$ pair,
exchanged by $\Omega$; the \Dfour-brane is constructed by imposing Dirichlet 
conditions in 5 dimensions, and adding a fermion degree of freedom to the
worldsheet boundary. The objects are tachyon-free even when the 32
background \Dbnine-branes are taken into account, and hence are fully
stable.

However, such objects can be destabilized in situations where they
encounter D5-branes. These are not present in the ten-dimensional
background, but however may arise in compactifications of the theory (see
\cite{adsau} for examples), where they may be required for RR tadpole
cancellation. Noticing that the gauge group on D5-branes in orthogonal,
the analysis of the \Dthree/D5 or \Dfour/D5 systems is analogous to that
in previous sections. Specifically, strings in mixed sectors produce
tachyons triggering the decay of the D3-brane (resp. D4-brane) into a
codimension two (resp. one) gauge lump in the D5-brane
group. Generalization of these statements to lower dimensional O$p$-planes
is straightforward.

\section{Non-BPS states and cohomology in M-theory}

In this section we report on the interesting structure of non-BPS states
of type IIA theory in the presence of an O6-plane. From the above
analysis, the theory in the absence of D6-branes (case on which we center
henceforth) contains fully stable non-BPS states constructed as non-BPS
\Dfour- and \Dfive-branes within the O6-plane. Namely, they can be
constructed as bundles on virtual D6-${\ov{{\rm D}6}}$ pairs, associated
to the non-trivial classes in ${\bf KO}(\IS^2)=\IZ_2$, 
${\bf KO}(\IS^1)=\IZ_2$. Our initial motivation was
to find the M-theory lifts of such states, which must exist given their
stability. In answering this question we will however be drawn to more
general, deeper, and in fact more interesting issues, related to the
interplay between K-theory in string theory and cohomology in M-theory,
and hence to the nature of RR and NS-NS charges in string theory.
Given this more general interest, it is convenient to start by reviewing
some general results on the topic.

The notion of K-theory as a classification tool for RR charges in string
theory is natural, since the objects charged under RR fields, D-branes, 
are naturally endowed with gauge bundles. Branes of M-theory, however, are
not, and K-theory does not seem to arise in M-theory in the same natural
way. Until a more complete framework is provided (or until it is 
convincingly shown that it is not required), the simplest proposal is that
charge under the M-theory $p$-forms is classified by cohomology. This
proposal has passed a first non-trivial check in the beautiful computation
in \cite{dmw}, where, for a large class of configurations, the zero mode
piece of the partition function for RR fields of type IIA on a manifold
${\bf X}_{10}$, topologically classified by K-theory, was recovered from
the zero mode piece of the partition function for $p$-form fields of
M-theory on ${\bf X}_{10}\times \IS^1$, as classified by cohomology. 

It is important to emphasize that in the configurations under study in
\cite{dmw}, in the string theory side the set of charges provided by K-theory 
is a {\em subset} of those provided by cohomology. Namely, there exist
cohomology classes which do not correspond to the Chern character of any
bundle, or K-theory class of bundles. In the M-theory computation, the
starting set of charges is given by cohomology, but the (properly defined)
$CGG$ phase in M-theory leads, upon reduction, to `projecting out' the
cohomology classes not corresponding to K-theory classes of the IIA string
configuration.

\medskip

It is natural to try to extend our understanding of the relation between
K-theory in string theory and its M-theory lift. There are several possible 
avenues to do so, and we would like to argue that the study of IIA theory
in the presence of an O6-plane explores some. First, it would be interesting 
to derive K-theory from M-theory on spacetimes which are $\IS^1$ bundles
over a 10d spacetime, rather than a Cartesian product. A simple example of
this type is provided by the M-theory lift of the O6-plane, which corresponds 
to an Atiyah-Hitchin manifold. A computation along the lines of \cite{dmw}
may be possible, but is beyond the scope of this note. Second, it is
natural to wonder about the M-theory origin of stable non-BPS states in
IIA theory, with charge classified by K-theory but not by cohomology. In
Section 5.3 we will describe one such state in IIA theory with an O6-plane, 
and show its M-theory lift corresponds to and M-brane with torsion
cohomological charge. Finally, it would be interesting to explore the role of 
NS-NS charged objects in the classification of topological charges in
string theory. In Section 5.2 we will show that lifting to M-theory
sometimes leads to topological equivalences of torsion NS-NS and RR
charges, suggesting the existence of some underlying structure unifying
their description. 

Our analysis is far from systematic, in particular given the uncertainty
of the correct framework to carry it out. We expect that our observations
concerning the O6-plane are helpful in finding it.

\subsection{Non-BPS states in IIA theory with an O6-plane}

Let us consider IIA theory with a negatively charged O6-plane, and no
D6-branes. As argued above, the theory contains certain non-BPS $\IZ_2$
charged states, constructed from \Dfour- and \Dfive-branes within the
O6-plane. A unified way to describe these and other states is to consider
branes wrapping non-trivial cycles in the transverse space $\IR^3/\IZ_2$. 
Since we are interested in compact cycles, states are classified by the
(co)homology of $\IRP$, given by (see e.g. \cite{hankol})
\beqa
\begin{array}{cccc}
H^0(\IRP,\IZ)=\IZ   \quad & H^0(\IRP,\ItZ)=0 \quad
& H_0(\IRP,\IZ)=\IZ   \quad & H_0(\IRP,\ItZ)=\IZ_2 \\
H^1(\IRP,\IZ)=0   \quad & H^1(\IRP,\ItZ)=\IZ_2 \quad
& H_1(\IRP,\IZ)=\IZ_2   \quad & H_1(\IRP,\ItZ)=0 \\
H^2(\IRP,\IZ)=\IZ_2   \quad & H^2(\IRP,\ItZ)=\IZ \quad
& H_2(\IRP,\IZ)=0   \quad & H_2(\IRP,\ItZ)=\IZ 
\end{array}\nonumber
\label{cohomah}
\eeqa
where $\ItZ$ is the twisted bundle of integers, whose cohomology
classifies periods of forms which change sign in moving around the
non-contractible path in $\IRP$.

Two comments are in order: First, D$(p+n)$-branes wrapped on $n$-cycles in
projective spaces have been extensively used \cite{sugi,flux} to classify
the possible fluxes giving different kinds of O$p$-planes (see e.g.
\cite{witbar,orienti, hankol,sugi,flux}). Here we are interested in
wrapping D$(q+n)$-branes 
on $n$-cycles to obtain $q$-brane states within O$p$-planes, for $q<p$.
Second, the classification given by homology should be refined by using
K-theory, as done for fluxes in \cite{flux}. We will not attempt to do so
in the present discussion, so our analysis may miss certain subtle 
features (like correlations between charges) \footnote{v2. We would like
to thank O.~Bergman for pointing out that the 1-brane charges below do not
T-dualize to charges of stable objects in type I theory, hence should not
correspond to conserved K-theory charges in the O6-plane (despite being
non-trivial in homology). Therefore a disappearance of homology classes
in going to K-theory seems to be at work, as in \cite{dmw,flux}.}.

The orientifold projection flips the sign of the IIA RR 1-, 5- and
9-forms, and of the NS-NS 2-form, hence D0-, D4-, D8-branes and fundamental 
strings (denoted F1's) may wrap cycles with non-trivial twisted
cohomology. The RR 3- and 7-forms, and the NS-NS 6-form are invariant,
hence D2-, D6- and NS5-brane wrappings are classified by usual homology
classes.

This construction provides an alternative description for some of the
states we know. For instance, a D4-brane wrapped on  the non-trivial class in
$H_0(\IRP,\ItZ)=\IZ_2$ corresponds to the \Dfour-brane directly
constructed above. Similarly, a D6-brane wrapped on the non-trivial class
in $H_1(\IRP,\IZ)=\IZ_2$ would correspond to the \Dfive-brane. We would
like to emphasize that the $\IZ_2$ charges of these states define
non-trivial cohomology classes, as opposed to the \Dseven- and
\Deight-branes in Sections 2, 3. This difference does not contradict
T-duality (which would apply in a compact context), this being related to
the fact that T-duality in general acts not within cohomology but in
derived categories (roughly speaking, K-theory) \cite{horisharpe}.

Instead of providing a complete classification, we would like to center on
$\IZ_2$ charged objects with tractable M-theory lifts. So we restrict our
attention for instance to: 

\begin{itemize}

\item a `RR' 4-brane obtained by wrapping a D4-brane on
the non-trivial class in $H_0(\IRP,\ItZ)=\IZ_2$

\item a `NS-NS' 4-brane given by an NS5-brane on the
non-trivial class in $H_1(\IRP,\IZ)=\IZ_2$.

\item a `RR' 1-brane obtained by wrapping a D2-brane on
the non-trivial class in $H_1(\IRP,\IZ)=\IZ_2$,

\item a `NS-NS' 1-brane given by an F1 on the
non-trivial class in $H_0(\IRP,\ItZ)=\IZ_2$

\end{itemize}
The M-theory lift of these states is discussed in next subsection.

\subsection{Non-BPS states in M-theory and NSNS/RR charge transmutation}

\begin{figure}
\begin{center}
\centering
\epsfysize=4cm
\leavevmode
\epsfbox{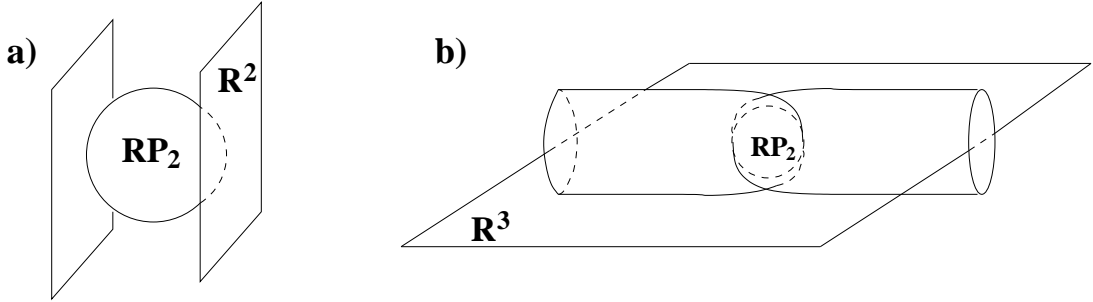}
\end{center}  
\caption[]{\small Schematic picture of $\IXAH$, a) as a topological space,
b) when endowed with a hyperkahler metric.} 
\label{ah}   
\end{figure}

The M-theory lift of the background O6-plane is given \cite{sw} by an
Atiyah-Hitchin manifold $\IXAH$, introduced in \cite{ah} (see \cite{hanpio} 
for diverse applications in string theory). To describe the topology of
$\IXAH$, consider an $\IS^1$ bundle over $\IS^2$, with Euler class $-4$,
modded out by a $\IZ_2$ action exchanging antipodal points in $\IS^2$ and
flipping the coordinate in $\IS^1$. This gives a circle bundle over
$\IRP$. $\IXAH$ is topologically an $\IR^2$ bundle over $\IRP$, where the
subbundle given by the angular $\IS^1$ is the above one \cite{sw}.
This space is schematically depicted in figure \ref{ah}a.

When $\IXAH$ is endowed with a hyperkahler metric, the $\IS^1$ fiber 
asymptotes to a constant radius, while the $\IRP$ grows along the
radial direction in the 2-plane fiber. The geometry is roughly that of a
$\IS^1$ fibration over $\IR^3$, modded out by $\IZ_2$, as depicted in
figure \ref{ah}b. M-theory on $\IXAH$ reduces, in the limit of small fiber
radius, to type IIA theory with an O6-plane (and without D6-branes). The
$\IRP$ within $\IXAH$ coincides with the $\IRP$ within $\IR^3/\IZ_2$ used 
above in the IIA configuration.

This information is enough to discuss the M-brane wrapping possibilities.
In fact, for the purpose of discussing the lifts of the above states we
just need to know that, since $\IXAH$ is homotopic to $\IRP$,
$H_1(\IXAH,\IZ)=\IZ_2$ with generator e.g. the non-contractible 1-cycle in
$\IRP$.

This leads to an interesting result. Since a D4-brane lifts to an
M5-brane wrapped on a 1-cycle, it follows that both the NS-NS and the RR
4-branes above correspond to an M5-brane wrapped on the non-trivial class
in $H_1(\IRP,\IZ)=\IZ_2$. This nicely reproduces the $\IZ_2$ charge of the
objects, but implies a more dramatic consequence. In fact, since there is
only one available 1-cycle, the NS-NS and RR 4-branes have an identical
M-theory lift, and hence are topologically indistinguishable. This is
sharp contrast with the string theory viewpoint, where they would seem to
excite different kinds of fields, NS-NS and RR respectively. However,
notice that the fields excited are pure torsion, so there is no
contradiction with familiar intuition about conservation of NS-NS and RR
charge. Similarly, the NS-NS and the RR 1-brane correspond to a M2-brane
on a 1-cycle.

\begin{figure}
\begin{center}
\centering
\epsfysize=4cm
\leavevmode
\epsfbox{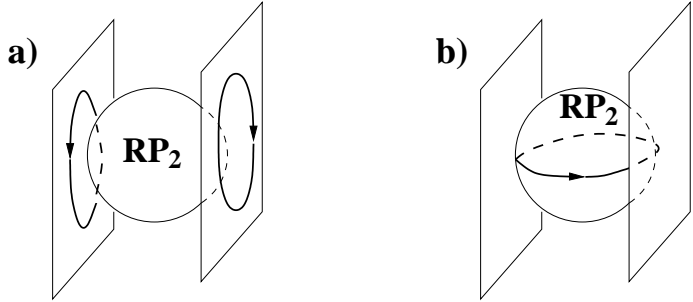}
\end{center}  
\caption[]{\small M-theory lift of the RR and NS-NS 4-branes as M5-brane
in non-trivial 1-cycles in $\IXAH$.}
\label{transmut}   
\end{figure}   

The situation is perhaps more intuitively grasped as follows. The RR
D4-brane is nothing but the familiar \Dfour-brane within the O6-plane,
analogous to the \Dseven-brane within the O9-plane in section 2. As such
one may understand it as a D4-${\ov{{\rm D}4}}$ pair exchanged by the
orientifold action. The pair can be moved away from the O6-plane in a
$\IZ_2$ symmetric fashion, leading to a very simple M-theory lift. It
corresponds to two M5-branes wrapped with opposite orientations on the
$\IS^1$ fiber of $\IXAH$, and sitting at opposite points on the base
$\IR^3$. The topology of the resulting cycle is shown in figure
\ref{transmut}a. The corresponding 1-cycle is a representative of the
non-trivial class in $H_1(\IRP,\IZ)=\IZ_2$. 

On the other hand, the NS-NS 4-brane is a NS5-brane wrapped on the
non-contractible 1-cycle in the $\IRP$ within $\IR^3/\IZ_2$. It lifts in
the obvious fashion to an M5-brane wrapped on the non-trivial cycle in the
$\IRP$ within $\IXAH$, providing just another representative of the
same non-trivial class in $H_1(\IRP,\IZ)=\IZ_2$. This cycle is shown in
figure \ref{transmut}b. 

It is not difficult to directly show that the two 1-cycles are in fact
homologous, and therefore the above two states carry the same charge.
Hence, a single homology class in M-theory allows to (un)wind torsion
objects in the M-theory circle, smoothly interpolating between branes with
torsion charge under NS-NS and RR fields in the weakly coupled IIA string
theory limit.

This result suggests the existence of an underlying framework in string
theory, describing the NS-NS and RR topological sectors and avoiding
redundancies like the one above. Such framework must go beyond the
familiar K-theory, which is not suitable to incorporate NS-NS charge.
It is remarkable that M-theory cohomology achieves this unified
description, and we hope that it should reduce to a sensible formalism in
the weakly coupled string theory regime.

\subsection{Non-BPS states in M-theory and cohomology}

Another issue we would like to address is the M-theory origin of string theory 
states with RR charge classified by K-theory but not by cohomology. The
O6-plane configuration provides us with a simple string background with
such charges. The question is hence how M-theory manages to incorporate
these charges, i.e. is M-theory cohomology still enough to take them into
account or do they require a more sophisticated framework?
In the following we show one example where M-theory cohomology is still
sufficient to take these states into account. The question of whether this
remains true for more general states and configurations remains open, and
clearly deserves further study.

The state we would like to study is not in the class considered above. In
intuitive terms, objects with too non-trivial transverse space tend to 
carry RR charges described by cohomology. We therefore try to construct
our purely K-theoretic state as extended in directions transverse to the
O6-plane. 

Consider a pair of (euclidean) D2-, \Dbtwo-branes, wrapped on
the three dimensions transverse to the O6-plane. This system is similar
(and related by T-duality) to a D${\widehat{(-1)}}$-brane on an O9-plane.
In fact, the orientifold projection exchanges the D2- and the
\Dbtwo-branes, removing their tachyon. More accurately, the tachyon is
projected out in the intersection with the O6-plane, but survives in the
bulk of the D2-${\ov{{\rm D}2}}$ pair. Hence, it may trigger asymptotic
annihilation of the objects, but not a complete one, since the state
carries non-trivial K-theory $\IZ_2$ charge. In this section, however, we
shall not be interested in the stability of the objects, rather in their
non-trivial charges. The $\IZ_2$ charge of the above object is easily seen
not to correspond to any cohomology class. If the RR $\IZ_2$ charge were
cohomological, the fact that asymptotic annihilation is possible indicates
that the cohomology class has compact support and should correspond to a
D-brane wrapped on a compact cycle in $\IRP$. However, there is no wrapped
D-brane candidate for it (for instance, a (euclidean) D0-brane cannot wrap
a 1-cycle, since $H_1(\IRP,\ItZ)$ is trivial), hence the $\IZ_2$ charge is
not cohomological.

\begin{figure}
\begin{center}
\centering
\epsfysize=4cm
\leavevmode
\epsfbox{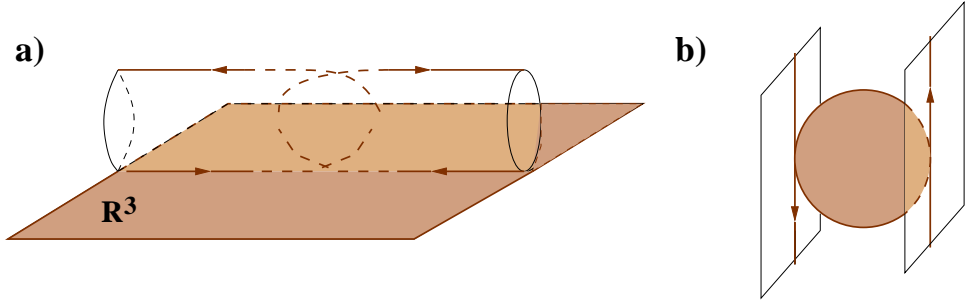}
\end{center}  
\caption[]{\small Schematic picture of the M-theory $\IZ_2$ charged
configuration of an M2-brane in a torsion 3-cycle in $\IXAH$. }
\label{kth}   
\end{figure}   

This state has an interesting and tractable M-theory lift. To describe it,
we use a different description of $\IXAH$. As discussed in \cite{ah,sw},
in a suitable complex structure $\IXAH$ can be described by the equation
\beqa
y^2=x^2v-4x
\label{eqah}
\eeqa
In order to compare with the geometry in the string theory configuration, it 
is convenient to describe its asymptotic behaviour by
\beqa
st=u^{-4}
\label{eqahtwo}
\eeqa
modded out by the $\IZ_2$ action $s \leftrightarrow t$, $u\to -u$.
Expression (\ref{eqah}) is obtained by defining the invariant variables
\beqa
y=(s-t)u \quad, \quad x=s+t+2u^{-2} \quad , \quad v=u^2. 
\label{identif}
\eeqa
The fact that the change of
variables is singular means that the description (\ref{eqahtwo}) is valid
only asymptotically, for large $u$. In this regime, the exponential
corrections in the core of $\IXAH$ can be ignored, and the geometry
reduces to a Taub-NUT space with `wrong sign' charge equal to $-4$. The
$\IZ_2$ action corresponds to the $\IZ_2$ reflection implied by the
O6-plane (roughly speaking, one may take $u=x^8+ix^9$, and $s/t\simeq
e^{x^7+ix^{10}}$). It is also interesting to observe that the $\IC^*$
action in $\IXAH$, $(x,y,v)\to (\lambda^2 x, \lambda y, \lambda^{-2} v)$,
whose $U(1)$ piece is a subgroup of the $SO(3)$ rotational symmetry
\cite{sw}, arises from a $\IC^*$ action $(s,t,u) \to (\lambda^2 s,
\lambda^2 t, \lambda^{-1} u)$, whose $U(1)$ subgroup is a rotation in the
string theory background. Also, the $\IC^*$ action $(s,t,u)\to(\lambda
s, \lambda^{-1} t, u)$, associated to the $U(1)$ rotation in the
asymptotic circle in `Taub-NUT' does not descend to $\IXAH$, which does
not have isometries in the circle fiber.

\medskip

From the viewpoint of M-theory, the D2-${\ov{{\rm D}2}}$ configuration
corresponds, far from the core in $\IXAH$, to an M2-brane wrapped on
oppositely oriented pieces of 3-cycle, located at definite positions in
the M-theory circle. In the description (\ref{eqahtwo}), one such 3-cycle
is given by the real condition
\beqa
s/t={\ov s}/{\ov t}
\label{three}
\eeqa
It describes a 3-cycle which spans the $\IR^3$ base of the `Taub-NUT', and is 
invariant under the $\IZ_2$ action. Moreover, it
intersects twice the asymptotic M-theory circle (given by the $U(1)$ orbit
$(s,t)\to (e^{i\alpha}s,e^{-i\alpha}t)$), hence the M2-brane looks like
a D2-${\ov{{\rm D}2}}$ pair in the string theory limit. A schematic
picture of these asymptotics is given in figure \ref{kth}a.

The two pieces of 3-cycle are accidentally located at opposite points in
the asymptotic $\IS^1$. However, there is no obstruction to approaching
them in a $\IZ_2$ invariant fashion, and one may be tempted to claim that
full annihilation follows. In fact, asymptotic annihilation is possible
(as explained above from the string theory viewpoint), but the twisting of
the geometry in the core of $\IXAH$ makes full annihilation impossible. In
order to see that, we need to extend the 3-cycle (\ref{three}) to the
interior of $\IXAH$. A suitable 3-cycle in (\ref{eqah}), asymptoting to
(\ref{three}), is given by
\beqa
y^2 \,v= {\ov y}^2\,{\ov v}
\label{threetwo}
\eeqa
The 3-cycle (\ref{threetwo}) is moreover invariant under the $\IC^*$
related to the $SO(3)$ rotational symmetry of the configuration. An
oversimplified picture of this cycle is shown in figure \ref{kth}b. 

In terms of variables in the asymptotic `Taub-NUT', this reads
\beqa
s/t + t/s = {\ov s}/{\ov t}+{\ov t}/{\ov s}
\eeqa
For $|s|$ much larger that $|t|$, this reduces to $s/t\in\IR$, and
viceversa, leading to correct asymptotic behaviour. However, in the near
core region, the 3-cycle (\ref{threetwo}) ends up winding around the $\IS^1$
fiber in $\IXAH$. This wrapping provides the M2-brane a non-trivial
charge, which prevents annihilation in the near core region.

Hence, the 3-cycle is non-trivial, and can be seen to be 2-torsion (two
M2-branes on the 3-cycle can annihilate to nothing by rotating them in the
$\IR^2$ fibers over $\IRP$ in $\IXAH$, more manifest in figure \ref{kth}b). 
Hence the purely K-theoretical RR charge for the string theory non-BPS
state arises from a purely cohomological charge in M-theory.

We conclude the discussion by mentioning that, after asymptotic annihilation 
in the M-theory configuration, we are left with an M2-brane wrapped on a
compact 3-cycle in $\IXAH$. This should be given by the $\IS^1$ fibration
over $\IRP$ mentioned at the beginning of Section 5.2. In the string theory 
limit the M2-brane therefore reduces to a fundamental string wrapped on
the $\IRP$ within $\IR^3/\IZ_2$ (this is possible since 
$H_2(\IRP,\ItZ)=\IZ_2$). Hence, this provides another example of
transmutation between RR and NS-NS torsion charges. Interestingly enough,
the charge in this case is not cohomological (but purely K-theoretical)
when interpreted as a RR torsion charge, but is cohomological when
interpreted as NS-NS charge.

\medskip

The states we have obtained are new non-BPS states in M-theory
\footnote{See e.g. \cite{mth} for other approaches to non-BPS physics in
M-theory}. As
discussed, they extend the results in \cite{dmw} on the relation between
cohomology in M-theory and K-theory in string theory. We expect further
study of more involved configurations to sharpen this correspondence. For
instance, type IIA theory with a positively charged O6-plane admits
diverse stable non-BPS states. Their M-theory lift is most intriguing,
since the O6$^+$-plane lifts to a mysterious frozen $D_4$ singularity
\cite{landlop}. Such states would presumably shed light into the nature of
M-theory brane charges in such background (see \cite{sav} for a recent
discussion). Clearly, the task of understanding the nature of the M-theory
3-form, the framework that classifies its topological sectors, and its
relation to NS-NS, RR charges and to K-theory in string theory will
require much more work. We hope our results are helpful in this quest.

\bigskip

\centerline{\bf Acknowledgements}

We would like to thank R.~Hern\'andez, K.~Landsteiner, E.~L\'opez,
M.~Gaberdiel, M.~Garc\'{\i}a P\'erez and K.~S.~Narain for useful
discussions. A.~M.~U. thanks  the Lawrence Berkeley Laboratory for
hospitality during completion of this work, and M.~Gonz\'alez for kind
encouragement and support. O. L.-B. thanks the TH-division at CERN for
hospitality and Hugo Garc\'{\i}a-Compe\'an for great support and very
useful suggestions. The work of O.~L.-B. is supported by a doctoral
CONACyT fellowship (M\'exico) under the program ``Estancias en el
extranjero para becarios nacionales (beca-cr\'edito mixta), 2000-2001'',
119267.

\bigskip

\appendix

\section{Appendix: Susy restoration in a type IIB orientifold}

In this subsection we show that understanding of the decay of the
\Dseven-brane can improve our understanding of certain compactifications 
of type I theory. In particular certain seemingly non-supersymmetric
compactifications involving \Dseven-branes have restored supersymmetry
once the \Dseven-brane tachyons relax to their minimum. 

A toy version would be to consider type I on $\IT^2$ with one 
\Dseven-brane {\em and} toron wilson lines turned on the D9-branes. The
system breaks supersymmetry, but the \Dseven-brane is unstable against
decay to a second toron which `unwinds' the first. The final model is the
standard toroidal compactification of type I, with unbroken supersymmetry.

The model we present is more involved, and certainly more interesting. In
particular, after the decay of the \Dseven-branes, the final configuration
is supersymmetric, but has a non-trivial distribution of $\IZ_2$ charges,
arising from different sources, which nevertheless add up to zero (as
required from K-theory RR charge cancellation \cite{probe}).

Our starting point is the $\Omega$ orientifold of type IIB compactified on
$\IT^4/\IZ_2$, constructed in \cite{bisagn,gp} \footnote{We hope no
confusion arises between the $\IZ_2$ orbifold group and the $\IZ_2$
K-theory charges.}. The model contains 32
D9-branes, and 32 D5-branes, as counted in the covering space. We consider
an initial configuration with trivial Wilson lines on the D9-branes, leading 
to a $U(16)$ gauge group, and D5-branes e.g. distributed among the 16 
$\IZ_2$ fixed points, yielding a product gauge group with total rank 16.

An important observation, as discussed in \cite{blpssw}, is that there are
certain constraints on the distribution of D5-branes among the $\IZ_2$
fixed points. In particular, for any four fixed points lying in a 2-plane,
the total number of D5-branes at such points must be $0$ $mod$ $4$.
In \cite{probe}, this constraint was shown to arise from cancellation of
the K-theory $\IZ_2$ charges, by showing that configurations not
satisfying the condition in \cite{blpssw} lead to global gauge anomalies
on the world-volume of D5-brane probes. 

This argument suggests that configurations violating the condition in
\cite{blpssw} may be rendered consistent by introducing type I 
\Dseven-branes cancelling the K-theory $\IZ_2$ charge. In the following we
present such an example, designed so that the set of \Dseven-branes
required is relatively simple. Consider a square $\IT^4$ with radii
$R_i$, and denote $N_{m_1,m_2,m_3,m_4}$ the number of D5-branes sitting at
the $\IZ_2$ fixed point with coordinates $x_i=m_i \pi R_i$ (hence,
$m_i=0,1$). Consider distributing 20 D5-branes as
\beqa
\begin{array}{cccccccc}
N_{0000}=2 & \quad ; \quad  N_{0100}=0 & \quad ; \quad  
N_{1000}=0 & \quad ; \quad  N_{0100}=0 \\
N_{0001}=0 & \quad ; \quad  N_{0101}=2 & \quad ; \quad  
N_{1001}=2 & \quad ; \quad  N_{0101}=2 \\
N_{0010}=0 & \quad ; \quad  N_{0110}=2 & \quad ; \quad  
N_{1010}=2 & \quad ; \quad  N_{0110}=2 \\
N_{0011}=0 & \quad ; \quad  N_{0111}=2 & \quad ; \quad  
N_{1011}=2 & \quad ; \quad  N_{0111}=2
\end{array}
\label{distri}
\eeqa
while the remaining 12 are located e.g. in the bulk in a $\IZ_2$ invariant
fashion. The distribution (\ref{distri}) violates the condition in
\cite{blpssw}, for instance along the 2-plane defined by $(x_1,x_2)=(0,0)$.
The non-cancellation of the $\IZ_2$ charge is reflected as a global gauge
anomaly on a D5-brane probe located at $(x_1,x_2)=(0,0)$, which contains
an odd number of $SU(2)$ doublet Weyl fermions \cite{probe}. 

The full K-theory charge may be cancelled by introducing one \Dseven-brane
located at a generic point in the $(x_1,x_2)$ plane and wrapped on 
$(x_3,x_4)$, and one \Dseven-brane at a generic point in the
$(x_3,x_4)$ plane and wrapped on $(x_1,x_2)$ \footnote{Notice
that these \Dseven-branes in $\IT^4/\IZ_2$ are orbifold invariant pairs of
\Dseven-branes in the double covering space.}. Any D5-brane probe on which
the background D5-branes induce a $\IZ_2$ anomaly intersects exactly one
\Dseven-brane, inducing a cancelling anomaly. Conversely, D5-brane probes
without anomaly from the background D5-branes do not get any fermions from
the \Dseven-branes. 

The resulting configuration is hence consistent, and seemingly
non-supersymmetric due to the \Dseven-branes. An amusing feature is that
supersymmetry is broken due to the necessity of cancelling the
(K-theoretical piece of) RR tadpoles, in a spirit similar to models
with branes and antibranes, or non-BPS branes, in the literature
\cite{sugimoto,adsau,abgru}.

However, the configuration is not stable, and we should study the final
state once the \Dseven-brane tachyons relax to their minimum.
Since the \Dseven-configuration is relatively simple, so is the answer.
The final, tachyon-free and stable, configuration has the same
distribution of D5-branes among the fixed points, but it has non-trivial
Wilson lines turned on the D9-branes. Their gauge group is $U(16)$, and
the Wilson lines $\gamma_{i}$ along $x_i$ can be taken
\beqa
\gamma_1= \diag(1,-1,-1;1,1,1;\id_{10}) & \quad ; \quad &
\gamma_3= \diag(1,1,1;1,-1,-1;\id_{10}) \\
\gamma_2= \diag(-1,1,-1;1,1,1;\id_{10}) & \quad ; \quad &
\gamma_4= \diag(1,1,1;-1,1,-1;\id_{10}) 
\eeqa
They correspond to two independent torons in the $\IT^2$'s transverse to
the original \Dseven-branes. Notice that, regarded in the parent $SO(32)$,
the Wilson lines (say) $\gamma_1$, $\gamma_2$ define {\em two} torons
exchanged by the orbifold group. This simply reflects that a single
\Dseven-brane in the quotient can be regarded as two \Dseven-branes in
$\IT^4$, exchanged by the orbifold group.

It is again possible to check that the K-theory $\IZ_2$ charge of the
configuration properly cancels, for instance by introducing D5-brane
probes wrapped on different 2-planes. Any such probe on which the
background D5-branes induce an anomaly, contains a compensating anomaly
from fermions in the 59+95 sectors.

Notice that the final configuration still violates the condition in
\cite{blpssw}. However, such condition was derived only in the absence of
D9-brane Wilson lines, and does not directly apply to our final model.
Instead, we have checked the consistency of the configuration by the
introduction of D-brane probes, as suggested in \cite{probe}. Our analysis
hence shows how to generalize the condition, and find new models consistent 
with K-theory charge cancellation. In the final model cancellation of the
$\IZ_2$ charge is quite non-trivial, with two kinds of sources (the
distribution of background D5-branes, and the toron D9-brane Wilson
lines) cancelling each other.

The final configuration is supersymmetric. Restoration of supersymmetry
after tachyon condensation was not obvious from the initial model. Hence
this example illustrates the importance of treating the \Dseven-brane
decay properly. We hope the remarks in this paper help in applying this
understanding in less academic examples.

\end{document}